\documentclass[12pt]{article}
\usepackage{amsmath, amssymb}
\usepackage{amssymb}
\usepackage{pstcol,epic,epsfig}
\begin{document}
\date{}
\title{$SU(2)$ Symmetry and Degeneracy From SUSY QM of a Neutron in the
Magnetic Field of a Linear Current}

\author{D Mart\'{\i}nez$^1$, V D Granados$^1$ and R D Mota$^2$}

\maketitle

\begin{minipage}{0.9\textwidth}
\small
$^{1}$ Escuela Superior de F\'{\i}sica y Matem\'aticas,
Instituto Polit\'ecnico Nacional, 
Ed. 9, Unidad Profesional Adolfo L\'opez Mateos, 07738 M\'exico D F, M\'exico.\\
$^{2}$ Unidad Profesional 
Interdisciplinaria de  Ingenier\'{\i}a 
y Tecnolog\'{\i}as Avanzadas, IPN.
Av. Instituto Polit\'ecnico Nacional 2580, Col. La Laguna Ticom\'an, 
Delegaci\'on Gustavo A. Madero, 
07340 M\'exico D. F., M\'exico.\\
\end{minipage}
\rm

E-mail: dmartinezs77@yahoo.com.mx (D Mart\'{\i}nez),\\ rdmotae@yahoo.com.mx 

\begin{abstract}
From SUSY ladder operators in momentum space of a neutron in the magnetic field of a linear current, we construct $2\times2$ matrix operators that together with the $z$-component of the angular momentum satisfy the $su(2)$ Lie algebra. We use this fact to explain the degeneracy of the energy spectrum.\\
\linebreak
\end{abstract}

\section{Introduction}
SUSY QM is the modern way to study solvable and perturbative systems as is extensively shown in Refs.[1-5]. Recently, it has been shown that for some solvable systems in more than one dimension, the SUSY operators can be used to explain the degeneracy of the spectrum [6-10].

The existence of magnetically bound states of a neutron embedded in the magnetic field of a filamentary current has been demonstrated experimentally by Schmiedmayer \cite{SCHMIED} and Vestergaard et. al. \cite{VESTE}. In the recent years this problem became very important from the experimental point of view. It is considered the basis for the so-called neutron magnetic storage, for the ultra cold neutrons physics and for the generation of radio frequency radiation \cite{IGNA}.

Pron'ko and Stroganov \cite{PRONKO} were the pioneers in studying this system analytically in momentum space. Even though they explained the degeneracy of the system, the origin of the symmetry generators is not given.
Bl\"umel and Dietrich \cite{BLUMEL01,BLUMEL02} studied the problem and found the energy spectrum and the eigenfunctions in configuration space. However, the works [14-16] do not show the supersymmetry of the problem. In \cite{VORONIN} Voronin solved the problem in momentum representation from SUSY QM. By expanding the spinor wave function in Fourier series, and after some transformations, Voronin obtained an exact supersymmetric form for each component of the expansion. He also found the energy spectrum and the eigenfunctions of the system. We note that the theoretical description developed in \cite{VESTE}, and more recently in \cite{LIMA}, was based upon the concept of SUSY QM for multicomponent wave functions in configuration space.
     
In this Letter we study the problem in momentum space. We construct its symmetry group generators from SUSY QM and explain its degeneracy of the energy spectrum.

Using polar coordinates we obtain an exact supersymmetric pair for the complete radial components of the spinor wave function. The second-order differential equation for each radial component is obtained, which is identified with the first P\"oschl-Teller equation. By using the SUSY QM approach [19,20], wwe get the ladder operators. Generalizing the supersymmetric operators to matrix operators of two variables, it is shown that these operators together with $\hat J_z$ satisfy the $so(3)$ or $su(2)$ Lie algebra. Using angular momentum theory, we find the energy spectrum, and show that these matrix operators allow us to find the set of states correspondng to a given energy level.

The magnetic field of an infinite straight wire carring a current $I$ located along the $z$-axis
\begin{equation}
\vec B=\frac{\mu_0 I}{2\pi}\frac{(-y,x,0)}{(x^2+y^2)}.
\end{equation}
Thus, the Hamiltonian of a neutron of mass $M=1$ and magnetic moment $\mu\vec\sigma$ interacting with $\vec B$ is 
\begin{equation}
\hat {\cal H}=\frac{\hat p^2}{2}+\mu{\vec\sigma}\cdot{\vec B}=\frac{\hat p^2}{2}+G\frac{-y\sigma_x+x\sigma_y}{(x^2+y^2)}
\end{equation}
where $G$ is a constant defined as
\begin{equation}
G=\frac{\mu \mu_0 I}{2\pi}.
\end{equation}
and we have set $c=\hbar=1$.  
Considering the translational symmetry along the $z$-axis, the two component wave function of the system can be written as
\begin{equation}
\Psi_{k} (\vec r)=\frac{1}{\sqrt L}e^{2\pi ikz/L}
\begin{pmatrix}
\psi_1(x,y)\cr\psi_2(x,y)
\end{pmatrix},
\label{psiconfig}
\end{equation}
where $k=0,1,2,3,...$ As a consequence of the free motion along the $z-$axis, we get a two dimensional problem. 

Thus, the Schr\"odinger equation can be written as a system of two differential equations
\begin{equation}
\frac{\hat p_x^2+\hat p_y^2}{2}\psi_1(x,y) +i 
G\frac{x-iy}{x^2+y^2}\psi_2(x,y)=\tilde E\psi_1(x,y),
\label{s1a}
\end{equation}
\begin{equation}
\frac{\hat p_x^2+\hat p_y^2}{2}\psi_2(x,y) -iG 
\frac{x+iy}{x^2+y^2}\psi_1(x,y)=\tilde E\psi_2(x,y),
\label{s1b}
\end{equation}
where $\tilde E\equiv E-\frac{2\pi k^2}{L^2}$.  
 
It is easy to show that $\left[\hat {\cal H},\hat J_z\right]=0$, where $\hat J_z=\hat L_z+\hat s_z$ is the $z$-component of the total angular momentum, therefore, using polar coodinates $(\rho,varphi)$, (\ref{psiconfig}) can be written as  
\begin{equation}
\Psi_{k\;j_z} (\vec r)=\frac{1}{\sqrt L}e^{2\pi ikz/L}
\begin{pmatrix}
\psi_1^{j_z}(\rho,\varphi)\cr\psi_2^{j_z}(\rho,\varphi)
\end{pmatrix}
=\frac{1}{\sqrt L}e^{2\pi ikz/L}
\begin{pmatrix}
f_1(\rho)e^{i(j_z-\frac{1}{2})\varphi}\cr f_2(\rho)e^{i(j_z+\frac{1}{2})\varphi}
\end{pmatrix},
\label{psic}
\end{equation}
where $j_z=\pm\frac{1}{2},\pm\frac{3}{2},...$ is the quantum number corresponding to the operator $\hat J_z$.

If we multiply Eqs. (\ref{s1a}) and (\ref{s1b}) by $(x+iy)$ and $(x-iy)$, respectively, we are able to write these equations in momentum space in the form
\begin{equation}
\left(\frac{\partial}{\partial p_x}+i\frac{\partial}{\partial p_y}\right)\alpha{\tilde\psi}_1^{j_z}(p_x,p_y)
=-G {\tilde\psi}_2^{j_z}(p_x,p_y),
\label{s2a}
\end{equation}
\begin{equation}
\left(\frac{\partial}{\partial p_x}-i\frac{\partial}{\partial p_y}\right)\alpha{\tilde \psi}_2^{j_z}(p_x,p_y)
=G{\tilde\psi}_1^{j_z}(p_x,p_y),
\label{s2b}
\end{equation}
where the variables $x$ and $y$ have been substituted by their corresponding operators in momentum space. In Eqs.(\ref{s2a}) and (\ref{s2b}) we have defined
\begin{equation}
\alpha\equiv \frac{p_x^2+p_y^2}{2}-\tilde E=\frac{p^2}{2}-\tilde E
\end{equation}
and $\tilde\psi_{a}(p_x,p_y)$, $a=1,2$ are the Fourier transforms of the functions $\psi_{a}(p_x,p_y)$ given by
\begin{equation}
\tilde\psi_{a}(p_x,p_y)=\frac{1}{2\pi}\int_{-\infty}^{\infty}\int_{-\infty}^{\infty}\psi_{a}(x,y)e^{-i(xp_x+yp_y)}dxdy
\label{psimom}
\end{equation}

Using polar coordinates in momentum space $(p,\theta)$, we write the cartesian components of the momentum $p$ as follows
\begin{equation}
p_x=p\cos\theta,
\end{equation}
\begin{equation}
p_y=p\sin\theta.
\end{equation}

Thus, Eq.(\ref{psimom}) can be written as
\begin{equation}
\tilde\psi_{i}^{j_z}(p,\theta)=\frac{1}{2\pi}\int_{0}^{\infty}\int_{0}^{2\pi}\psi_i^{j_z}(\rho,\varphi),
e^{-i(p\rho\cos(\theta-\varphi))}\rho d\rho d\varphi
\end{equation}
where $\psi_i^{j_z}(\rho,\varphi)$ are defined in Eq.(\ref{psic}). Therefore,
\begin{equation}
\tilde\psi_{i}^{j_z}(p,\theta)=i^{-(j_z\mp\frac{1}{2})}e^{i(j_z\mp\frac{1}{2})\theta}\int_{0}^{\infty}\rho f_{i}(\rho)J_{j_z\mp\frac{1}{2}}(p\rho)d\rho\equiv e^{i(j_z\mp\frac{1}{2})\theta}F^{j_z}_{i}(p),
\label{psim}
\end{equation}
where $J_{j_z\mp\frac{1}{2}}$ are the Bessel funtions of order $j_z\mp\frac{1}{2}$. The result given by equation (\ref{psim}) permits to write the spinor wave function as
\begin{equation}
{\tilde \Psi}_{j_z} (\vec {p})=
\begin{pmatrix}
e^{i(j_z-\frac{1}{2})\theta}F_1^{j_z}(p)\cr e^{i(j_z+\frac{1}{2})\theta}F_2^{j_z}(p)
\end{pmatrix},\label{spinor}
\end{equation}
which implies  the conservation of the operator $\hat J_z$ in momentum space, as was expected. Explicitly, we have
\begin{equation}
\hat J_z\tilde\Psi_{j_z}(\vec p)=
\begin{pmatrix}
-i\frac{\partial}{\partial\theta}+\frac{1}{2}&0\cr0&-i\frac{\partial}{\partial\theta}-\frac{1}{2}
\end{pmatrix}
\begin{pmatrix}
e^{i(j_z-\frac{1}{2})\theta}F_1^{j_z}(p)\cr e^{i(j_z+\frac{1}{2})\theta}F_2^{j_z}(p)
\end{pmatrix}
=j_z\tilde\Psi_{j_z}(\vec p)\label{jz}
\end{equation}

Therefore, Eqs. (\ref{s2a}) and (\ref{s2b}) are transformed to 
\begin{equation}
\left(\frac{d}{dp}-\frac{{j_z}-\frac{1}{2}}{p}\right)\alpha F_1^{j_z}(p)=-G F_2^{j_z}(p)
\label{s3a},
\end{equation}
\begin{equation}
\left(\frac{d}{dp}+\frac{{j_z}+\frac{1}{2}}{p}\right)\alpha F_2^{j_z}(p)=G F_1^{j_z}(p).
\label{s3b}
\end{equation}

If we perform the change of variables defined by
\begin{equation}
q=p(-2\tilde E)^{-1/2},
\end{equation}
\begin{equation}
Z^{j_z}_i=p^{1/2}(\alpha F_i^{j_z}),\label{Zj}
\end{equation}
now, Eqs. (\ref{s3a}) and (\ref{s3b}) can be expressed as
\begin{equation}
\left(\frac{d}{dq}-\frac{j_z}{q}\right)Z^{j_z}_1=-\frac{\lambda^{1/2}}{\frac{q^2}{\Lambda}+1}Z^{j_z}_2,
\label{s4a}
\end{equation}
\begin{equation}
\left(\frac{d}{dq}+\frac{j_z}{q}\right)Z^{j_z}_2=\frac{\lambda^{1/2}}{\frac{q^2}{\Lambda}+1}Z^{j_z}_1,
\label{s4b}
\end{equation}
respectively, with 
\begin{equation}
\lambda^{1/2}=\frac{2G}{\hbar(-2\tilde E)^{1/2}}.
\label{esp1}
\end{equation}

Now, setting $q=\tan{\beta\over2}$, we introduce the angular variable $\beta$, $0\le\beta\le\pi$. Thus, equations (\ref{s4a}) and (\ref{s4b}) are transformed to
\begin{equation}
\left(-\frac{d}{d\beta}+W(\beta)\right)Z^{j_z}_1=\epsilon^{1/2}Z_2^{j_z},
\label{s5a}
\end{equation}
\begin{equation}
\left(\frac{d}{d\beta}+W(\beta)\right)Z^{j_z}_2=\epsilon^{1/2}Z^{j_z}_1
\label{s5b},
\end{equation}
where 
\begin{equation}
\epsilon=\frac{\lambda}{4},
\label{espectro}
\end{equation} 
\begin{equation}
W(\beta)=\frac{j_z}{2}\left(\tan{\beta\over2}+\cot{\beta\over2}\right).\label{superpot}
\end{equation}
If we consider the following definitions,
\begin{equation}
\hat A^\dag=-\frac{d}{d\beta}+W(\beta),\hspace{4ex}\hat A=\frac{d}{d\beta}+W(\beta),
\end{equation}
Eqs.(\ref{s5a}) and (\ref{s5b}) can be written as
\begin{equation}
\hat A^\dag Z_1^{j_z}=\epsilon^{1/2}Z_2^{j_z},\label{s6a}
\end{equation}
\begin{equation}
\hat A Z_2^{j_z}=\epsilon^{1/2}Z_1^{j_z}.\label{s6b}
\end{equation}
We recognize that Eqs.(\ref{s5a}) and (\ref{s5b}) are written in the so-called exact supersymmetric form, being $W(\beta)$ the superpotential \cite{DUTT}. Operators $\hat A^\dag$ and $\hat A$ can be used to define the supersymmetric Hamiltonians $\hat H_{j_z:+}\equiv\hat A\hat A^\dag$ and $\hat H_{j_z:-}\equiv\hat A^\dag\hat A$. From Eq.(\ref{s6a}) and (\ref{s6b}) we can see that they operate on $Z_1^{j_z}$ and $Z_2^{j_z}$, respectively. Say
\begin{equation}
\hat H_{j_z:+}Z_1^{j_z}=\hat A\hat A^\dag Z_1^{j_z}=\epsilon Z_1^{j_z},\label{par+}
\end{equation} 
\begin{equation}
\hat H_{j_z:-}Z_2^{j_z}=\hat A^\dag\hat A Z_2^{j_z}=\epsilon Z_2^{j_z}.\label{par-}
\end{equation}
This means that the radial components of the spinor (\ref{spinor}), $Z_1^{j_z}$ and $Z_1^{j_z}$, are the eigenfunctions of the SUSY Hamiltonians $\hat H_{j_z:+}$ and $\hat H_{j_z:-}$, respectively. Notice that, from (\ref{par+}) and (\ref{par-}), the spectrum of the Hamiltonians $\hat H_{j_z:+}$ and $\hat H_{j_z:-}$ is the same $\epsilon\equiv\epsilon^{\pm}$, which implies a broken supesymmetry \cite{BAGCHI}.

It is mmediate to find the spectrum $\epsilon$ because we have identified (\ref{superpot}) as the superpotential of the problem. By using Table I of Ref.\cite{DUTT}, we find that the superpotential (\ref{superpot}) is a particular case of
\begin{equation}
W=A\tan\alpha x-B\cot\alpha x,
\end{equation}
which corresponds to the P\"oschl-Teller I potential, with spectrum
\begin{equation}
E_n^{(-)}=(A+B+2n\alpha)^2-(A+B)^2
\end{equation}
and the ground state
\begin{equation}
\psi_0^{(-)}=(\sin\alpha x)^{B/\alpha}(\cos\alpha x)^{A/\alpha}.
\end{equation}
For our case, if we set $A=-B=j_z/2$ and $\alpha=1/2$, then
\begin{equation}
\epsilon^{\pm}=n^2,\hspace{3ex}n=1,2,3,...,
\end{equation}
\begin{equation}
\psi_0^{(-)}=(\sin\alpha x)^{-j_z}(\cos\alpha x)^{j_z}.\label{psi0}
\end{equation}
Because, $j_z=\pm\frac{1}{2},\pm\frac{3}{2},...,$ the ground state (\ref{psi0}) is non-normalizable. It means that the state with $n=0$ in $\epsilon^-$ is abscent, which implies $\epsilon^-=\epsilon^+$, as we have moted after equation (\ref{par-}). This confirms that SUSY is broken \cite{BAGCHI}.
It is remarkable that we have used changes of variable analogous to those used by Voronin\cite{VORONIN} and obtained similar results. Nevertheless, the difference between our and that of Voronin is that we have not expanded the wave function in space. This allowed us to find the SUSY pair (\ref{s5a}) and (\ref{s5b}) for the complete radial wave funtion in momentum space.
The main results we will obtain in this Letter are based on the following fact: since the supersymmetric Hamiltonians $\hat H_{j_z:+}$ and $\hat H_{j_z:-}$ are differential operators of second order, each of them can be factorized in a different way to SUSY. Indeed, by the Infeld-Hull (IH) factorization method \cite{INFELD}.
Explicitly, Eqs. (\ref{par+}) and (\ref{par-}) are

\begin{equation}
\hat H_{j_z;\pm}\Phi_{j_z}^\pm(\beta)=\left( -\frac{d^2}{d\beta^2}+V_{j_z;s_z}(\beta)\right)\Phi_{j_z}^\pm(\beta)=\epsilon \Phi_{j_z}^\pm(\beta),
\label{pt}
\end{equation}
with
\begin{equation}
V_{j_z;\pm}(\beta)={1\over4}\Bigg[{({j_z}-s_z+{1\over 2})
({j_z}-s_z-{1\over 2})\over\sin^2{\beta\over2}}+{({j_z}+s_z+{1\over 2})({j_z}+s_z-{1\over 2})\over\cos^2{\beta\over2}}\Bigg],
\label{pt1}
\end{equation}
where we have defined $\Phi_{j_z}^+=Z_1^{j_z}(\beta)$ and $\Phi_{j_z}^+=Z_1^{j_z}(\beta)$. 
According to \cite{INFELD}, we find that the IH ladder operators corresponding to tje first P\"oschl-Teller potential (\ref{pt1}) are
\begin{equation}
\hat B_{j_z;\pm}(\beta)=-\frac{d}{d\beta}+k_\pm(\beta),\label{B}
\end{equation}
\begin{equation}
\hat B_{j_z;\pm}^\dag(\beta)=\frac{d}{d\beta}+k_\pm(\beta),\label{Bdag}
\end{equation}
where
\begin{equation}
k_\pm(\beta,j_\pm)=\left(\frac{j_z}{2}+\frac{1}{4}\mp\frac{1}{4}\right)\cot\frac{\beta}{2}-\left(\frac{j_z}{2}-\frac{1}{4}\pm\frac{1}{4}\right)\tan\frac{\beta}{2}.
\end{equation}

By straightforward calculation, it is easy to show that the IH ladder operators factorize the Hamiltonians $\hat H_{j_z;\pm}$ in the following way
\begin{align}
\hat B_{j_z;\pm}& (\beta)\hat B_{j_z;\pm}^\dag(\beta)\Phi_{j_z}^\pm(\beta)\nonumber\\
& =\left[\hat H_{j_z;\pm}-\left(j_z-\frac{1}{2}\right)^2\right]\Phi_{j_z}^\pm(\beta),
\label{a1}
\end{align}
\begin{align}
\hat B_{j_z+1;\pm}^\dag& (\beta)\hat B_{j_z+1;\pm}(\beta)\Phi_{j_z}^\pm(\beta)\nonumber\\
& =\left[\hat H_{j_z;\pm}-\left(j_z+\frac{1}{2}\right)^2\right]\Phi_{j_z}^\pm(\beta).
\label{a2}
\end{align}
Eqs. (\ref{a1}) and (\ref{a2}) can be written in the matrix form as
\begin{align}
& \begin{pmatrix}
\hat B_{j_z;+}(\beta)\hat B_{j_z;+}^\dag(\beta)&0\cr0&\hat B_{j_z;-}(\beta)\hat B_{j_z;-}^\dag(\beta)
\end{pmatrix}
\begin{pmatrix}
\Phi_{j_z}^+(\beta)\cr\Phi_{j_z}^-(\beta)
\end{pmatrix}\nonumber\\
& =\begin{pmatrix}
\hat H_{j_z;+}-\left(j_z-\frac{1}{2}\right)^2&0\cr0&\hat H_{j_z;-}-\left(j_z-\frac{1}{2}\right)^2
\end{pmatrix}
\begin{pmatrix}
\Phi_{j_z}^+(\beta)\cr\Phi_{j_z}^-(\beta)
\end{pmatrix},\label{b1}
\end{align}
\begin{align}
&\begin{pmatrix}
\hat B_{j_z+1;+}^\dag(\beta)\hat B_{j_z+1;+}(\beta)&0\cr0&\hat B_{j_z+1;-}^\dag(\beta)\hat B_{j_z+1;-}(\beta)
\end{pmatrix}
\begin{pmatrix}\Phi_{j_z}^+(\beta)\cr\Phi_{j_z}^-(\beta)
\end{pmatrix}\nonumber\\
&=
\begin{pmatrix}
\hat H_{j_z;+}-\left(j_z+\frac{1}{2}\right)^2&0\cr0&\hat H_{j_z;-}-\left(j_z+\frac{1}{2}\right)^2
\end{pmatrix}
\begin{pmatrix}
\Phi_{j_z}^+(\beta)\cr\Phi_{j_z}^-(\beta)
\end{pmatrix}.\label{b2}
\end{align}
We have found the IH one variable ladder operators $\hat B_{j_z;\pm}$ and $\hat B_{j_z;\pm}^\dag$which factorize the corresponding Hamiltonians $\hat H_{j_z;\pm}$. Notice that these operators depend on the quantum number $j_z$. To consturct the symmetries of the system, we generalize the IH ladder operators and $\hat H_{j_z;\pm}$ changing $j_z$ by its corresponding operator in the variable $\theta$. This is achieved as follows. From Eqs. (\ref{par+}) and (\ref{par-}), and the definitons after Eq. (\ref{pt1}), $\hat H_{j_z;+}$ and $\hat H_{j_z;-}$ act on the upper ($\Phi_{j_z}^+(\beta)$) and the lower ($\Phi_{j_z}^-(\beta)$) radial components of the spinor wave function, respectively. According to Eq. (\ref{jz}), the quantum number $j_z$ must be changed by $-i\frac{\partial}{\partial\theta}+\frac{1}{2}$ if ti is within an operator acting on $\Phi_{j_z}^+(\beta)$, and by $-i\frac{\partial}{\partial\theta}-\frac{1}{2}$ if ti is within an operator acting on $\Phi_{j_z}^-(\beta)$. Therefore, operators defined in Eqs.(\ref{B}) and (\ref{Bdag}) can be generalized to
\begin{align}
\hat B_\pm=& e^{i\theta}\Big\lbrace-\frac{\partial}{\partial\beta}+\frac{1}{2}\Big[\left(-i\frac{\partial}{\partial\theta}-\frac{1}{2}\right)\cot\frac{\beta}{2}\nonumber\\
&-\left(-i\frac{\partial}{\partial\theta}\pm1-\frac{1}{2}\right)\tan\frac{\beta}{2}\Big]\Big\rbrace,\label{genB}
\end{align}
\begin{align}
\hat B_\pm^\dag=& e^{-i\theta}\Big\lbrace-\frac{\partial}{\partial\beta}+\frac{1}{2}\Big[\left(-i\frac{\partial}{\partial\theta}+\frac{1}{2}\right)\cot\frac{\beta}{2}\nonumber\\
&-\left(-i\frac{\partial}{\partial\theta}\pm1+\frac{1}{2}\right)\tan\frac{\beta}{2}\Big]\Big\rbrace,\label{genBdag}
\end{align}
and the Hamiltonians $\hat H_{j_z;\pm}$ are generalized to
\begin{align}
\hat H_\pm=-\frac{\partial^2}{\partial\beta^2}& +\frac{1}{4}\Big[\left(-i\frac{\partial}{\partial\theta}\pm1+\frac{1}{2}\right)\left(-i\frac{\partial}{\partial\theta}\pm1-\frac{1}{2}\right)\sec^2\frac{\beta}{2}\nonumber\\
& +\left(-i\frac{\partial}{\partial\theta}-\frac{1}{2}\right)\left(-i\frac{\partial}{\partial\theta}+\frac{1}{2}\right)\csc^2\frac{\beta}{2}\Big].\label{genH}
\end{align}
Notice that in Eqs.(\ref{genB}) and (\ref{genBdag}) we have added a phase factor because these operators act on the omplete spinor components $e^{i(j_z\pm\frac{1}{2})\Phi^\pm_{j_z}}(\beta)$.
From Eqs.(\ref{genB}), (\ref{genBdag}) and (\ref{genH}) we define the $2\times2$ matrix operators
\begin{equation}
\hat J_+=
\begin{pmatrix}
\hat B_+&0\cr0\cr&\hat B_-
\end{pmatrix},
\hspace{3ex}
\hat J_-=
\begin{pmatrix}
\hat B_+^\dag&0\cr0\cr&\hat B_-^\dag
\end{pmatrix}\label{J+-}
\end{equation}
and
\begin{equation}
\hat H=
\begin{pmatrix}
\hat H_+&0\cr0\cr&\hat H_-
\end{pmatrix}.\label{Hmat}
\end{equation}
Also, we define the spinor wave function
\begin{equation}
Z_{j_z}(\beta,\theta)=
\begin{pmatrix}
e^{i(j_z-\frac{1}{2})}\Phi_{j_z}^+(\beta)\cr e^{i(j_z+\frac{1}{2})}\Phi_{j_z}^-(\beta)
\end{pmatrix}.\label{spinor}
\end{equation}
The definitons given by Eqs. (\ref{J+-}), (\ref{Hmat}) and (\ref{spinor}) allow us to generalize the matrix equations (\ref{b1}) and (\ref{b2}) as
\begin{equation}
\hat J_{+}\hat J_{-}Z_{j_z}(\beta,\theta)=\left[\hat H-\left(\hat J_z+\frac{1}{2}\right)^2\right]Z_{j_z}(\beta,\theta).
\label{c1}
\end{equation}
\begin{equation}
\hat J_{-}\hat J_{+}Z_{j_z}(\beta,\theta)=\left[\hat H-\left(\hat J_z-\frac{1}{2}\right)^2\right]Z_{j_z}(\beta,\theta).
\label{c2}
\end{equation}

The last two equations can be written in compact form as
\begin{equation}
\hat J_{\pm}\hat J_{\mp}Z_{j_z}(\beta,\theta)=\left[\hat H-\left(\hat J_z\pm\frac{1}{2}\right)^2\right]Z_{j_z}(\beta,\theta).
\label{c3}
\end{equation}

These equations, together with
\begin{equation}
\hat J_z Z_{j_z}(\beta,\theta)=\left(\hat L_z+\hat s_z\right)Z_{j_z}(\beta,\theta)=j_z Z_{j_z}(\beta,\theta),
\end{equation}
lead us to the commutation relations
\begin{equation}
\left[\hat J_+,\hat J_-\right] Z_{j_z}(\beta,\theta)=2\hat J_z Z_{j_z}(\beta,\theta)
\end{equation}
\begin{equation}
\left[\hat J_z,\hat J_\pm\right] Z_{j_z}(\beta,\theta)=\pm\hat J_{\pm} Z_{j_z}(\beta,\theta),
\label{degen}
\end{equation}
which is the Lie algebra of one of the groups $SO(3)$ or $SU(2)$. By using this angular momentum algebra, we find that 
\begin{equation}
\left[{\hat J}^2,{\hat J}_+\right]=\left[{\hat J}^2,{\hat J}_-\right]=\left[{\hat J}^2,{\hat J}_z\right]=0
\label{comrules}
\end{equation}
and
\begin{equation}
\hat J_\mp\hat J_\pm={\hat J}^2-\hat J_z\left(\hat J_z\pm1\right).
\label{ang1}
\end{equation}
Also, from Eqs. (\ref{c3}) and (\ref{comrules}) it follows that
\begin{equation}
[\hat J^2,\hat H]=0.
\end{equation}
Thus, from the IH factorization method we have constructed the operators $\hat J_\pm$. These operators together with $\hat J_z$ allowed us to obtain the Casimir operator $\hat J^2$, which commutes with $\hat J_z$ and with the Hamiltonian $\hat H$. Therefore, $\hat H$, $\hat J^2$ and $\hat J_z$ are a compkete set set of commuting operators which have simultaneus eigenfunctions. This fact permits to introduce the principal quantum number $j$ which satisfies
\begin{equation}
{\hat J}^2 Z_{j,j_z}(\beta,\theta)=j(j+1) Z_{j,j_z}(\beta,\theta)
\end{equation}
\begin{equation}
{\hat J}_z Z_{j,j_z}(\beta,\theta)=j_z Z_{j,j_z}(\beta,\theta),
\end{equation}
with $-j\le j_z\le j$.
Substitution of (\ref{ang1}) into Eq. (\ref{c3}) leads us to
\begin{equation}
\left[{\hat J}^2-\hat J_z\left(\hat J_z\pm1\right)\right] Z_{j,j_z}(\beta,\theta)
=\left[\hat H-\left(\hat J_z\pm{1 \over 2}\right)^2\right] Z_{j,j_z}(\beta,\theta).
\label{c4}
\end{equation}
From each of these equations we find that the spectrum of the operator $\hat H$ is given by
\begin{equation}
\epsilon_j =\left(j+{1\over 2} \right)^2,
\label{esp3}
\end{equation}
which is in agreement with  that reported in the literature \cite{LANGE}. Using Eqs. (\ref{esp1}), (\ref{espectro}) and (\ref{esp3}), we find that the spectrum of $\hat{\cal H}$ is
\begin{equation}
{\tilde E}_j=-\frac{G^2}{2\left(j+\frac{1}{2}\right)^2}.
\label{finalespectro}
\end{equation}

From the commutation relations (\ref{degen}), it is immediate to show that
\begin{equation}
\hat J_z\hat J_\pm Z_{j,j_z}=
\left(j_z\pm1\right)\hat J_\pm Z_{j,j_z},
\end{equation}
which implies that  $\hat J_\pm Z_{j,j_z}$ is an eigenvector of $\hat J_z $ with eigenvalue $j_z\pm 1$. This means that $\hat J_\pm Z_{j,j_z}\propto Z_{j,j_z\pm 1}$. In fact, by taking the same convection of the angular momentum theory, we get\begin{equation}
\hat J_+ Z_{j,j_z}(\beta,\theta)=\sqrt{j(j+1)-{j_z}(j_z+1)} Z_{j,j_z+1}(\beta,\theta)\label{4a}
\end{equation}
\begin{equation}
\hat J_- Z_{j,j_z+1}(\beta,\theta)=\sqrt{j(j+1)-{j_z}(j_z+1)} Z_{j,j_z}(\beta,\theta)\label{4b}.
\end{equation}
Notice that Eq. (\ref{espectro}) was obtained by using SUSY QM, whereas Eq. (\ref{esp3}) was obtained using angular momentum theory. Regarding that $j_z=\pm\frac{1}{2},\pm\frac{3}{2},\pm\frac{5}{2},...$, then, $j=\frac{1}{2},\frac{3}{2},\frac{5}{2},...$, thus, Eq. (\ref{espectro}) and (\ref{esp3}) reduce to the same energy spectrum. Since each of these equations does not depend on the quantum number $j_z$, the energy spectrum is degenerated. The restriction $-j\le j_z\le j$ implies that the degeneracy of the energy spectrum $\epsilon_j$ is $2j+1$. This number is not easy to find by other way to that given here. A very important result of our treatment is that, because $j=\frac{1}{2},\frac{3}{2},\frac{5}{2},...$, the symmetry gropu of the problem is $SU(2)$, the covering group of the rotation gropu $SO(3)$ \cite{SCHIFF}.

Eqs. (\ref{4a}) and (\ref{4b}) show that operators $\hat J_\pm$ can be used to obtain the set of degenerate states belonging to the enery level $\epsilon_j$.

We note that the energy spectrum, Eq. (\ref{finalespectro}), as well as its degeneracy are equal to thoseof the two-dimensional hydrogen atom [24,25]. Moreover, it has been shown that, as a limit case, the degenerate wave wunctions and the spectrum of the problem we have considered here, reduce those of the two-dimensional hydrogen atom \cite{BLUMEL01}, whose symmetry group is $SO(3)$ \cite{CISNEROS}.

We have obtained the symmetries of the system in momentum space bacause it is a two-dimensional system. This procedure os very difficult to perform for systems in more than two dimensions \cite{MOTA03}.

As a final remark, the procedure described here can be applied to the same problem in configuration space. This is the subject of a forthcoming report.

\section*{Acknowledgments}
We thank the referee by his suggestions which were very fruitful to the final version of the Letter.

This work was partially supported by CONACyT-M\'exico, COFAA-IPN, EDD-IPN, PIFI, CGPI-IPN project numbers 20050193, 2005855 and 20050182.

\end{document}